\documentclass[aps,pre%
,twocolumn%
,preprintnumbers,
,amsfonts,showpacs,amssymb]{revtex4}
\usepackage{graphics}
\usepackage{graphicx}
\newcommand{\jtit}[1]{{}}
\newcommand{\linksshift}[1]{\mbox{}\hspace*{-#1}}
\renewcommand{\linksshift}[1]{{}}
\begin{document}
\title{Non-Gaussian fluctuations arising from finite populations:
Exact results 
\\for the evolutionary Moran process}
\author{Jens Christian Claussen}
\email{claussen@theo-physik.uni-kiel.de}
\author{Arne Traulsen}
\email{traulsen@theo-physik.uni-kiel.de} 
 \affiliation{
\mbox{Institut f{\"u}r Theoretische Physik und
Astrophysik, Christian-Albrechts Universit{\"a}t,
Olshausenstra{\ss}e 40, 24098 Kiel, Germany}
}
\date{September 24, 2004}
\preprint{Phys.\ Rev.\ E 71, 025101(R) (2005) --  Rapid Communications}
\begin{abstract}
The appropriate description of fluctuations within the 
framework of evolutionary game theory is a fundamental 
unsolved problem in the case of finite populations. 
The Moran process recently introduced into this context
in Nowak {\em et al.}, [Nature (London) {\bf 428}, 646 (2004)] 
defines a promising standard model of evolutionary game theory
in finite populations for which analytical results are accessible.
In this paper, we derive the stationary distribution of the 
Moran process population dynamics for arbitrary $2\times{}2$ games 
for the finite size case. We show that a nonvanishing background 
fitness can be transformed to the vanishing case by rescaling the 
payoff matrix. In contrast to the common approach to mimic finite-size 
fluctuations by Gaussian distributed noise, the finite-size fluctuations 
can deviate significantly from a Gaussian distribution.
\end{abstract}
\pacs{%
\vspace*{-0.6em}
02.50.Le,
05.45.-a,
87.23.-n,
89.65.-s
\vspace*{-1.2em}
\hfill
DOI: 10.1103/PhysRevE.71.025101
}
\maketitle

Theoretical studies of coevolutionary dynamics usually 
assume infinite populations, as the replicator dynamics 
\cite{Hofbauer1998,Smith1981}
or the
Lotka-Volterra 
equations 
\cite{lotka,volterra}.
The limit of infinite populations 
leading to deterministic
differential
equations
is an
idealization motivated mainly by 
mathematical convenience.
Only in few cases the population will be large enough 
to justify the 
assumption of infinite populations.

In finite populations, crucial differences can appear. 
Population
states that cannot be invaded by a small fraction of 
mutants in infinite 
population,
so-called
Evolutionary Stable Strategies \cite{Hofbauer1998},
can be invaded by a single mutant 
\cite{Nowak2004b}.
In addition, a certain
inherent stochasticity is always present in finite populations. 
In multipopulation
interactions, 
such fluctuations can possibly be exploited 
\cite{TRS}.
In this paper, we quantify the inherent 
fluctuations arising from finite 
populations. 
As a starting point, we 
investigate
the classical Moran process 
\cite{Moran1962}
that was recently transfered to frequency dependent selection 
\cite{Nowak2004b,c_taylor}.
In a Moran process, 
in each time step
one agent is replicated and one agent is eliminated. 
Thus the total size of 
the population is strictly conserved.
This process can be considered as a standard model 
for game dynamics in
finite 
populations. 
Although a strictly fixed population size
will be fulfilled only in systems with
hard resource limitations,
e.g.\ a fixed number of academic positions,
it
is a widely common default, 
especially in spatial games
\cite{lim02,fort04,johnson99,TCa,togashi04}.
From a systematic point of view,
the dynamics within this process and the nature of the 
fluctuations have to be understood before a generalization
to variable population sizes on solid grounds is possible.

In \cite{TCb} we have shown that 
the Moran process introduced in \cite{Nowak2004b}
can be derived as a mean-field approximation
of the finite population game dynamics. 
In mean-field theories of evolutionary game theory
\cite{Hauert2002,TS,Szabo2002,laessig_euro}
not only the spatial degrees of freedom
are neglected;
but
the limit of infinite populations 
also
implies a transition from
a stochastic system to a deterministic equation of motion.
While the average effect of mutations can often be lumped in a 
deterministic term \cite{TS,eriksson_lindgren_jtb2},
different ways to incorporate external stochasticity 
have been proposed, e.g.\
by a Langevin term of Gaussian distributed noise
\cite{Foster1990,cabrales,TRS}
or stochastic payoffs \cite{eriksson_lindgren_gecco01}.
Consequently, one could also approximate
the intrinsic noise of the finite system 
by Gaussian noise reintroduced into the continuum equations.
But {\sl a priori} it is not clear, in which
situation this approximation is justified.
Especially in small populations, the inherent 
stochasticity
may significantly exceed any external noise.
In a finite-round Prisoner's Dilemma game, the broadness of the
distribution of cooperators recently was found to
promote cooperation \cite{mcnamara}.
Further, the distribution decay of fluctuations is known to 
be of substantial impact both 
in genetic evolutionary dynamics \cite{iwasa_genetics04}
and in evolutionary optimization \cite{boettcher_percus01}.

\paragraph*{\sl To clarify} the nature of inherent fluctuations 
of evolutionary dynamics
in a Moran process is the scope of this paper.
We quantify the deviations from the mean value by explicitly calculating 
the stationary distribution of strategies for 
general $2{\times}2$ games 
and provide a transformation for the case of 
nonvanishing background fitness.
The process is illustrated with two qualitatively representative 
kinds of games, and the exact solution, also for the
more realistic situation of a nonvanishing background fitness,
is provided.

\paragraph*{Moran evolution dynamics in $2{\times}2$ games.---}
We consider a finite population of N agents of two different types,
A and B, interacting in a game with the payoff matrix
\begin{equation}
P=\left( \begin{array}{cc} a & b \\ c & d \\ \end{array} \right).
\end{equation}
Each agent
interacts with a certain number of randomly chosen partners. 
The A individual $s$ obtains the fitness
\begin{equation}
\pi^A_s = 1-w+w\frac{n^A_s a + n^B_s b}{n^A_s + n^B_s},
\end{equation}
where $n^A_s$ ($n^B_s$) is the number of interactions with
A (B) individuals. $0 \leq w \leq 1$ measures the contribution of the game
to the fitness, $1-w$ is the background fitness. An equivalent equation
holds for B agents. Occasionally, 
the payoff of a randomly chosen
individual $s$ is compared
with the payoff of another randomly chosen
agent $u$. 
With probability $\pi_s/(\pi_s+\pi_u)$, a copy
of agent $s$ replaces agent $u$. With probability $\pi_u/(\pi_s+\pi_u)$,
agent $s$ is replaced by a copy of $u$. The probability that an agent 
reproduces is hence proportional to its payoff. The
payoff depends on the type of the individual and on the kind of its 
interactions. This approach is frequently used in simulations of 
multiagent systems 
\cite{vicsek95,schweitzer_holyst,helbing97,wong04}, 
genetic algorithms \cite{Holland1975, Mitchell1996}, and evolutionary 
game theory \cite{Hauert2002}.

The averaged dynamics of this model can be computed
from a mean-field theory \cite{TCb}. 
If every agent interacts with a 
representative sample of the population, the average 
payoff of $A$ and $B$ individuals will be,
 respectively,  
\begin{eqnarray}
\label{1payoff}
\pi^A(i) & = & 1 - w + w \frac{a(i-1)+b(N-i)}{N-1} \\ \nonumber
\pi^B(i) & = & 1 - w + w \frac{c \, i+d(N-1-i)}{N-1},
\end{eqnarray}
where $i$ is the number of $A$ individuals. We explicitly
excluded self interactions. 
An individual is selected for reproduction with
a probability proportional to its payoff, as described above. 
It replaces an individual that is chosen at random. 
This reduces the process to a Moran process \cite{Moran1962},
which was recently transfered to a game theoretic context 
\cite{Nowak2004b,c_taylor}.
The corresponding mean-field dynamics is 
given by a Markov process with the transition probabilities \cite{TCb}
\begin{eqnarray}
\label{transmat}
& T_{i\rightarrow i+1} & =  \frac{\pi^A(i)\,i      }{\pi^A(i) i + \pi^B(i)\,(N-i)}   \frac{N-i}{N} \\ \nonumber
& T_{i \rightarrow i-1}  & = \frac{\pi^B(i)\, (N-i)}{\pi^A(i) i + \pi^B(i)\,(N-i)} \frac{i}{N} \\ \nonumber
& T_{i \rightarrow i} &    = 1 - T_{i\rightarrow i+1} -T_{i \rightarrow i-1} .
\end{eqnarray}
All other transition probabilities are zero. 
The states $i=0$ and $i=N$ are 
absorbing, while the remaining states are transient. 
Conveniently, a small mutation can be introduced to
allow for an escape from the absorbing states
\cite{fudenberg_imhoff2004}.

\paragraph*{The general case of nonvanishing background fitness.} 
For a nonvanishing background fitness $1-w>0$ the 
transition properties obtained directly from 
Eqs.\ (\ref{1payoff}) and (\ref{transmat}) 
become quite lengthy.
A more elegant way
is to 
rescale the payoff matrix of a given $2{\times}2$ game
according to
\begin{eqnarray}
\left(\begin{array}{cc}
a^{\prime}&b^{\prime}\\
c^{\prime}&d^{\prime}
\end{array}\right)
=
\left(\begin{array}{cc}
1+(a-1)w~&~1+(b-1)w\\
1+(c-1)w~&~1+(d-1)w
\end{array}\right).
\label{wtransformation}
\end{eqnarray}
With this rescaled payoff matrix, a vanishing background fitness
can be assumed in (\ref{1payoff}) without loss of generality.

\begin{figure}[thbp]
\begin{center}
\includegraphics[width=60mm,angle=270]{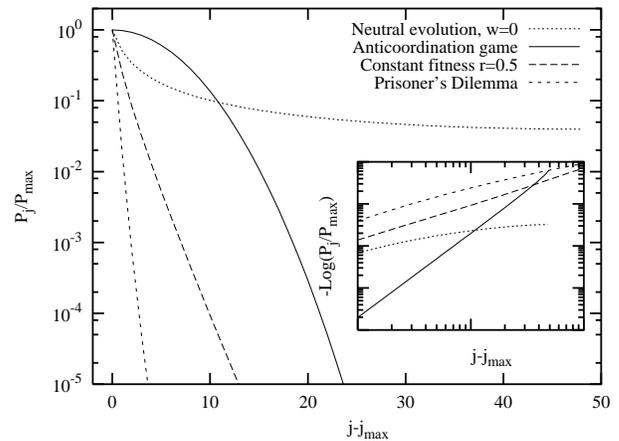}
\end{center}
\vspace*{-3ex}
\caption{Stationary probability distribution for 
different evolutionary dynamics depending on 
the distance to the maximum (N=100).
For comparison, also the slow decay 
for neutral evolution is shown. 
The decay of the distribution can be fitted
by a stretched exponential $\exp(-bx^\gamma)$ 
with $\gamma=2.06$ (anticoordination game),
$\gamma=0.87$ (constant fitness), 
and $\gamma=0.63$ (Prisoner's Dilemma).
The inset shows the same data where both axes 
are logarithmized, thus stretched exponentials
appear as straight lines.
The decay deviates significantly from a Gaussian 
distribution
for constant fitness and Prisoner's Dilemma,
corresponding to a random motion in an anharmonic potential.
}
\label{distribution}
\normalsize
\end{figure}

\paragraph*{Fluctuations around the average strategy:}
In order to quantify the deviations from the average
strategy of the system, we compute the stationary 
distribution 
$P_i$
for this system. 
We assume a small 
mutation probability $\mu$.   For $\mu \ll1$,
mutations affect the system only in the states that are
absorbing for $\mu =0$. 
In this case, the strategy distribution 
is generated only by the inherent stochasticity of the finite
population. 
The stationary probability can be computed in the
interior independently from the boundaries, the correct
normalization can then be found analyzing the transitions
from the boundaries to the interior, i.e. $P_0 \mu =P_1 T_{1 \rightarrow 0}$.

Let us first consider the 
{\sl neutral evolution}
limit of $w = 0$, where the fitness
is constant and independent of the type.
The payoffs are $\pi^A(i)=\pi^B(i)=1$. 
This implies 
\begin{equation}
T_{i\rightarrow i+1}=T_{i \rightarrow i-1} = \frac{i(N-i)}{N^2}.
\end{equation}
From $P_i \, T_{i\rightarrow i+1} = P_{i+1} \, T_{i+1\rightarrow i}$ we find 
in equilibrium for $0<i <N$
\begin{equation}
P_i \propto \frac{1}{(N-i)i}
\label{neutral_p}
\end{equation}
which has a minimum at $i=N/2$. The equilibrium distribution
arises from a neutral evolution of two types,
as known from population genetics \cite{Ewans1979}.

\paragraph*{Constant fitness.}
The simplest case for $w>0$ is the case of constant fitness,
i.e., $a=b<c=d=1$. The evolutionary dynamics drifts towards the 
type B, which has higher fitness. 
We find for the stationary probability
distribution 
($0<i<N-1$)
\begin{equation}
\frac{P_{i+1}}{P_i} = r \frac{r (i+1)+N-i-1}{r i + N-i} 
\frac{i}{i+1}
\frac{N-i}{N-i-1},
\label{thesimplest}
\end{equation}
where $r=1-w+w a<1$. 
Far from the borders (at $i=0,N$),
${P_{i+1}}/{P_i}$
converges to $r$ 
implying an exponential decay
of the stationary probability distribution.

\clearpage

\begin{figure}[thbp]
\begin{center}
\includegraphics[width=\columnwidth,angle=0]{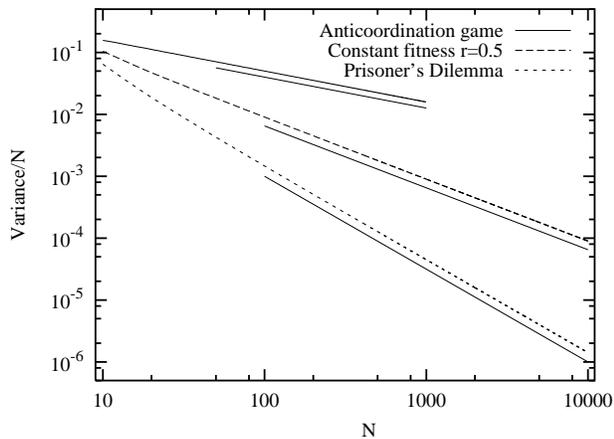}
\end{center}
\vspace*{-3ex}
\caption{Scaling of the variance, normalized by $N$, 
of the finite-size
fluctuations 
for anticoordination game (slope $-1/2$),
 constant fitness (slope $-1$), and
Prisoner's dilemma  (slope $-3/2$).   
For neutral evolution (not shown) the
variance increases faster than N.
}
\label{variance}
\normalsize
\end{figure}

\paragraph*{Internal Nash equilibrium.}
For frequency dependent fitness and $w>0$, the game can 
have an internal Nash equilibrium or an
equilibrium in one of the absorbing states.

As a simple example with an internal Nash equilibrium
we choose a simple ``anticoordination'' game with $w=1$,
\vspace*{-.6ex}
\begin{equation}
P=\left( \begin{array}{cc} 0 & 1 \\ 1 & 0 \\ \end{array} \right).
\vspace*{-.6ex}
\end{equation}
For the transition probabilities, we find  
\vspace*{-.6ex}
\begin{eqnarray}
T_{i\rightarrow i+1}&=&\frac{N-i}{2 N} \\ \nonumber
T_{i \rightarrow i-1}&=&\frac{i}{2N},
\vspace*{-.6ex}
\end{eqnarray}
which describes a random walk with a drift 
towards the deterministic fixed point
$i=N/2$. In equilibrium, we have $P_i \, T_{i\rightarrow i+1} = P_{i+1} \, T_{i+1\rightarrow i}$
for every $i$, which leads to
\vspace*{-.4ex}
\begin{equation}
P_{i+1} = P_0 \prod_{j=0}^{i} \frac{N-j}{j+1} =  P_0 \left( \begin{array}{c} N \\ i+1 \end{array}
\right),
\vspace*{-.6ex}
\end{equation}
where $P_0$ is determined by normalization.
$P_i$ is a binomial distribution 
around the equilibrium of the replicator dynamics
at $i=N/2$, 
$P_i = 2^{-N}$
$\!\big(${\small\footnotesize$\!\!\begin{array}{c} N \\ i
\end{array}\!\!$}$\big)$.

\paragraph*{Prisoner's Dilemma: Nash equilibrium at the border.}
The Prisoner's Dilemma \cite{Axelrod1984}
is a standard model,
where mutual cooperation leads to highest payoff
in the iterated game.
It is motivated by the situation where
two prisoners can reduce their time in prison
by witnessing the other's guilt (``defect'').
On the other hand, if both ``cooperate'' and refrain from blaming
the other,
 both receive
a reduction of punishment. 
This is described with parameters fulfilling $c>a>d>b$;
the dilemma situation originates from the temptation $c>a$,
defection yields a higher payoff if the opponent cooperates.
In its standard parameters,
the Prisoner's Dilemma is defined by
the payoff matrix
\vspace*{-1.6ex}
\begin{equation}
P=\left( \begin{array}{cc} 3 & 0 \\ 5 & 1 \\ \end{array} \right).
\vspace*{-1ex}
\end{equation}
which has a Nash equilibrium for mutual defection, i.e.\ $i=0$. 
As $b=0$, also state $i=1$ is absorbing for $w=1$ (two cooperators are
needed to promote cooperation).
Thus a small mutation rate $\mu$
has to be assumed also for $T_{1{\to}2}$.
Alternatively
one could assume $w<1$.
The transition probabilities are given by
\begin{eqnarray}
T_{i\rightarrow i+1} & = & \frac{3i-3}{-i^2-2i+3iN+N(N-1)} \frac{i(N-i)}{N} \\ \nonumber
T_{i \rightarrow i-1} & = & \frac{4i+N-1}{-i^2-2i+3iN+N(N-1)} \frac{i(N-i)}{N}.
\end{eqnarray}
From this, a
closed form of the probability distribution can
be derived (see below
for a derivation with arbitrary payoff matrix). 
A comparison between different stationary distributions 
is shown in Fig.\ \ref{distribution}.
The finite-size scaling of the variance is shown for the
same cases in Fig.\ \ref{variance}.

\paragraph*{Stationary Distribution for an arbitrary payoff matrix.} 
For the ratio of the transition probabilities between $i$ and $i+1$ 
we find with $w=1$,
cf.\ Eq.~(\ref{transmat}),
\begin{widetext}
\begin{eqnarray}
\nonumber
\!\!\!\!\!\!\!\!\!\!\!\!\!\!\! 
\frac{T_{i\rightarrow i+1}}
{T_{i+1 \rightarrow i}}
\!\!&\!=\!&\!\! 
\frac{\pi^{A}(i)
}{
i \pi^{A}(i) + (N-i)\pi^{B}(i)
}
\;
\frac{
(i+1) \pi^{A}(i+1) + (N-i-1)\pi^{B}(i+1)
}{
\pi^{B}(i+1)
}
\;
\frac{
i(N-i)
}{
(i+1)(N-i-1)
}
\\
\!\!\!\!\!\!\!\!\!\!\!\!\!\!\!\!\!\!\! 
\!\!&\!=\!&\!\!
\nonumber
\frac{a(i-1)+b(N-i)}{c(i+1)+d(N-i-2)}
\frac{
i(N-i) 
[ (i+1)^2(a-b-c+d)+(i+1)(-a+bN+cN+d-2dN) +N(N-1)d] 
}{
(i+1)(N-i-1)
[ i^2(a-b-c+d)+i(-a+bN+cN+d-2dN) +N(N-1)d] 
}
\!\!\!\!\!\!\!\!\!\!\!\!\!\!\!\!\!\!\!
\\
\!\!\!\!\!\!\!\!\!\!\!\!\!\!\!\!\!\!\! 
\!\!&\!=\!&\!\!
\frac{a-b}{c-d}
\;
\frac{i
-N_5
}
{i
-N_6
}
\;
\frac{i(N-i)
}{
(i+1)(N-i-1) 
}
\;
\frac{(i-N_1) (i-N_3)}{(i-N_2) (i-N_4)}.
\end{eqnarray}
Here $N_1 \cdots N_4$ are the 
roots of the quadratic expressions in $i$
and $N_5=\frac{a-bN}{a-b}, N_6=\frac{c+d(N-2)}{d-c}$.
We have excluded the special cases $a-b=0$, $c-d=0$
discussed above in (\ref{thesimplest})
and $(a-b)/(c-d)=1$, where some factors
do not depend on $i$ and part of the expression simplifies.
For $N-1 > k \geq j > 1$, the density of the stationary state 
can be solved explicitly
giving rising factorials (Pochhammer symbols), or equivalently,
quotients of Gamma functions,
\noindent
\begin{eqnarray}   
\linksshift{1.8cm}
\!\!\!
\!\!\!
\frac{P_k}{P_j}
\!\!\!
&=&
\!\!\!
\prod_{i=j}^{k-1}
\frac{T_{i\rightarrow i+1}}
{T_{i+1 \rightarrow i}}
=
\left(\frac{a-b}{c-d}\right)^{k-j}
\!\!\!
\cdot
\frac{
j (N-j)
}{
k (N-k)
}
\cdot
\frac{
\Gamma(k-N_5) 
\Gamma(j-N_6)
\Gamma(k-N_1) 
\Gamma(j-N_2) 
\Gamma(k-N_3)
\Gamma(j-N_4)    
}
{
\Gamma(j-N_5)
\Gamma(k-N_6)
\Gamma(j-N_1) 
\Gamma(k-N_2) 
\Gamma(j-N_3) 
\Gamma(k-N_4)
}
~~~~
\label{gammaformel}
\end{eqnarray}
\end{widetext}
which yields, after 
calculating $P_N/P_{N-1}$ and $P_1/P_0$ explicitly,
and after
normalization, the total density 
of the stationary state.
Equations (\ref{wtransformation}) and 
(\ref{gammaformel}) cover the general case of
$2{\times}2$ games including nonvanishing background
fitness.
The previously discussed examples
 are included as 
special cases.

{\sl\it To conclude,}
the distribution of the fluctuations around
a Nash equilibrium can be
nontrivially
broadened in 
realistic models of evolutionary game theory.
We analyzed the effect of
internal noise stemming from the
inherent evolutionary update fluctuations
in a finite population.
In general, internal noise
and externally imposed stochastic
forces can follow qualitatively different distributions.
In our paper, we concentrated on the important
case of a Moran process, which can be considered as
a standard model of evolutionary game dynamics in 
finite populations.
For the Moran process,
the effect of the finite size
of the population can be accessed directly.
Neglecting external noise, we have shown
that the stationary distribution of the 
Moran process of evolutionary $2{\times}2$ games
can be calculated analytically and 
yields different decay tails of the
distributions.
Depending on the payoff matrix 
and the location of the Nash equilibrium,
the finite size fluctuations may deviate significantly
from a Gaussian distribution.
\\
[2ex]
\indent 
We thank M.A.\ Nowak and S.\ Bornholdt for valuable comments 
on previous versions of the manuscript.
A.T.\ acknowledges support by the Studienstiftung des deutschen Volkes (German National Academic Foundation).
\bibliographystyle{plain}

\begin{thebibliography}{99}
\bibitem{Hofbauer1998} 		
J.\ Hofbauer and K.\ Sigmund,
{\em Evolutionary Games and Population Dynamics}
(Cambridge University Press,
Cambridge, England,
1998).


\bibitem{Smith1981}
J.M.\ Smith, 
{\em Evolution and the Theory of
Games}
(Cambridge University Press, 
Cambridge, England,
1982).




\bibitem{lotka}
A.J.\ Lotka,
J.\ Am.\ Chem.\ Soc.\ {\bf 42},1595
(1920).

\bibitem{volterra}
V.\ Volterra,
%
{\em  Variazioni e fluttuazioni del numero d'individui in specie animali conviventi, }
Mem.\ Acad.\ Lincei.\ (Roma) {\bf 2}, 31 --113
(1926).





\bibitem{Nowak2004b}	M. A. Nowak, A. Sasaki, C. Taylor, and D. Fudenberg,
			Nature (London) {\bf 428}, 646 (2004).


\bibitem{TRS}		A. Traulsen, T. R{\"o}hl, and H. G. Schuster, 
			Phys. Rev. Lett.  {\bf 93}, 028701 (2004).



\bibitem{Moran1962}	
P. A. P. Moran,
{\em The Statistical Processes of Evolutionary Theory},
(Clarendon, Oxford, UK, 1962).



\bibitem{c_taylor} 
C.\ Taylor, D.\ Fudenberg, A.\ Sasaki, and M.A.\ Nowak,
Bull. Math. Biol., 
{\bf  66},
1621 (2004).





\bibitem{lim02}
Y.F.\ Lim, K.\ Chen, and C.\ Jayaprakash,
\jtit{\sl Scale-invariant behavior in a spatial game of prisoner's dilemma,}
Phys. Rev. E {\bf 65}, 26134 (2002).

\bibitem{fort04}
H.\ Fort and S.\ Viola,
\jtit{\sl Self-organization in a simple model of adaptive agents playing
$2{\times}2$ games with arbitrary payoff matrices,}
Phys. Rev. E {\bf 69}, 36110 (2004).

\bibitem{johnson99}
{N.F.\ Johnson, P.M.\ Hui, R.\ Jonson, and T.S.\ Lo,}
\jtit{\sl Self-Organized segregation within an Evolving Population,}
Phys. Rev. Lett. {\bf 82}, 3360 (1999).



\bibitem{TCa} A.\ Traulsen and J.C.\ Claussen, 
\jtit{\sl Similarity based cooperation and spatial segregation,}
Phys. Rev. E
{\bf 70}, 046128 (2004).




\bibitem{togashi04}
{Y.\ Togashi and K.\ Kaneko,}
\jtit{\sl Molecular discreteness in reaction-diffusion systems yields steady states not seen in the continuum limit,}
Phys. Rev. E {\bf 70}, 020901(R) (2004).



\bibitem{TCb} A.\ Traulsen and J.C.\ Claussen,
cond-mat/0409655. 
--
\\
{\small\sl
Published as:
Arne Traulsen, Jens Christian Claussen and Christoph Hauert,
Coevolutionary Dynamics: From Finite to Infinite Populations,
Physical Review Letters 95, 238701 (2005).
}



\bibitem{TS}		
%
A.\ Traulsen and H.G.\ Schuster, Phys.\ Rev.\ E {\bf 68}, 046129 (2003).



\bibitem{Hauert2002} C.\ Hauert, Int.\ J.\ Bifurcation Chaos 
Appl.\ Sci.\ Eng.\ 
{\bf 12}, 1531 (2002).

\bibitem{Szabo2002} G.\ Szab{\'o} and C.\ Hauert, Phys.\ Rev.\ Lett.\ {\bf 89}, 118101 (2002).

\bibitem{laessig_euro}
M.\ L\"assig, F.\ Tria, and L.\ Peliti,
\jtit{\sl Evolutionary games and quasispecies,}
Europhys. Lett. {\bf 62}, 446--451 (2003).





\bibitem{Foster1990} 
D.\ Foster and P.\ Young,
\jtit{\sl Stochastic Evolutionary Game dynamics,}
Th.\ Pop.\ Biol.\
{\bf 38}%
, 219
(1990).

\bibitem{cabrales} 
A. Cabrales,
\jtit{\sl Stochastic Replicator Dynamics,}
Int.\ Econom.\ Rev.\
{\bf 41}%
, 451
(2000).


\bibitem{eriksson_lindgren_jtb2} 
A.\ Eriksson and K.\ Lindgren,
\jtit{\sl  Cooperation driven by mutations in multi-person Prisoner's Dilemma,}
J.\ Theor.\ Biol.
{\bf 232},
 399 
(2004).




\bibitem{eriksson_lindgren_gecco01}  
A.\ Eriksson and K.\ Lindgren,
\jtit{\sl Evolution of strategies in repeated stochastic games with full information of the payoff matrix,}
in: Proceedings of the Genetic and
Evolutionary Computation Conference GECCO 2001, edited by
H. Beyer et al., Morgan Kaufmann publications (San Francisco),
pp. 853--859 (2001).








\bibitem{mcnamara}
J.M.\ McNamara, Z.\ Barta, and A.I.\ Houston,
\jtit{\sl Variation in behaviour promotes cooperation in the Prisoner's Dilemma game,}
Nature (London) {\bf 428}, 745
 (2004).



\bibitem{iwasa_genetics04}
Y.\ Iwasa, F.\ Michor, and M.A.\ Nowak,
\jtit{\sl Stochastic Tunnels in Evolutionary Dynamics,}
Genetics {\bf 166}, 1571
 (2004).

\bibitem{boettcher_percus01}
{S.\ Boettcher and A.G.\ Percus,}
\jtit{\sl Optimization with Extremal Dynamics,}
Phys. Rev. Lett. {\bf 86}, 5211 (2001).



\bibitem{vicsek95} 
{T.\ Vicsek, A.\ Czirok, E.\ Ben Jacob, I.\ Cohen, and O.\ Shochet,}
\jtit{\sl Novel type of phase transitions in a system of self-driven particles,}
Phys. Rev. Lett. {\bf 75}, 1226
 (1995).

\bibitem{schweitzer_holyst}
{F.\ Schweitzer and J.A.\ Ho{\l}yst,}
\jtit{\sl Modelling Collective Opinion Formation by Means of Active Brownian Particles,}
Eur.\ Phys.\ J.\ B {\bf 15}, 723 
 (2000).
%

\bibitem{helbing97}
D.\ Helbing, F.\ Schweitzer, J.\ Keltsch, and P.\ Moln\'ar,
\jtit{\sl Active walker model for the formation of human and animal trail systems,}
Phys.\ Rev.\ E {\bf 56}, 2527
 (1997).




\bibitem{wong04}
K.Y.\ Michael Wong, S.W.\ Lim, and Z.\ Gao,
\jtit{\sl Dynamical mechanisms of adaptation in multiagent systems,}
Phys.\ Rev.\ E {\bf 70},
025103(R)  (2004).



\bibitem{Holland1975} J. Holland, 
{\em Adaptation in Natural and Artificial Systems},
(MIT Press, Cambridge, Mass., 1975).


\bibitem{Mitchell1996} 	
M. Mitchell, {\em An Introduction to Genetic Algorithms},
(MIT Press, Cambridge, Mass., 1996).



\bibitem{fudenberg_imhoff2004}
 D.\ Fudenberg and L.A.\ Imhoff,
Harvard Inst.\ Econ.\ Res.\ Discussion Paper 2050,
unpublished (2004).
Available online from
http://post.economics.harvard.edu/hier/




\bibitem{Ewans1979}
W.J.\ Ewens,
{\it Mathematical Population Genetics},
(Springer,
Berlin,
1979).



\bibitem{Axelrod1984}
R. Axelrod, {\em The Evolution of Cooperation},
(Basic Books, New York, 1984).
















\end{thebibliography}

\end{document}